\documentclass{aa}\usepackage{graphics,epsf}
\begin{document}
\thesaurus{06(06.13.1; 02.03.1; 02.13.1)}
\title{Torsional oscillations in the solar convection zone}
\author{Eurico Covas\thanks{e-mail: eoc@maths.qmw.ac.uk}\inst{1}
\and Reza Tavakol\thanks{e-mail: reza@maths.qmw.ac.uk}\inst{1}
\and David Moss\thanks{e-mail: moss@ma.man.ac.uk}\inst{2}
\and Andrew Tworkowski\thanks{e-mail: ast@maths.qmw.ac.uk}\inst{3}}
\institute{Astronomy Unit, School of Mathematical Sciences,
Queen Mary and Westfield College, Mile End Road, London E1 4NS, UK
\and Department of Mathematics, The University, Manchester M13 9PL, UK
\and MRC, School of Mathematical Sciences,
Queen Mary and Westfield College, Mile End Road, London E1 4NS, UK.
}
\date{Received ~~ ; accepted ~~ }
\offprints{\it E.\ Covas}
\maketitle
\markboth
{Covas {\em et al.}: Torsional oscillations in the solar convection zone}
{Covas {\em et al.}: Torsional oscillations in the solar convection zone}
\begin{abstract}
Recent analysis of the helioseismic
observations
indicate that the previously observed
surface torsional oscillations
extend significantly  downwards
into the solar convection zone.

In an attempt to understand these oscillations,
we study the nonlinear coupling between
the magnetic field and the solar differential rotation in the context
of a mean field dynamo model,
in which the nonlinearity is due to the action
of the azimuthal component of the Lorentz force of the
dynamo generated magnetic field on the solar
angular velocity.
The underlying zero order angular velocity is
chosen to be consistent with the most recent helioseismic data.

The model produces butterfly diagrams which are in qualitative agreement
with the observations.
It displays
torsional oscillations that penetrate
into the convection zone,
and which with time  migrate towards the equator.
The period of these oscillations is found to be
half that of the period of
the global magnetic fields. This is
compatible with the observed period
of the surface torsional oscillations.
Inside the convection zone,
this is a testable prediction
that is not ruled out by the
observations so far available.
\end{abstract}

\keywords{Sun: torsional oscillations; interior;
rotation; magnetic fields;  mean field dynamos.}

\section{Introduction}
An important feature of the
solar convection zone is the presence of
differential rotation in the form of
a decrease in angular
velocity from equator to the pole. This has been
observed both in the surface layers (e.g.\ Snodgrass
1984) and deeper in the convection zone, as inferred from helioseismic
measurements
(e.g.\ Thompson {\em et al.} 1996).
Furthermore, the differential rotation on the surface has
been observed to vary with time (e.g.\ Howard \& LaBonte 1980; Snodgrass, Howard, \& Webster 1985).
These so called torsional oscillations, which have periods
of about 11 years,
manifest themselves in the form of four alternating latitudinal bands
of slightly faster and slower than average zonal flows
which migrate towards the equator in about 22 years.
These oscillations
have also been confirmed by the
analysis of the helioseismic data from the Solar and Heliospheric
Observatory (SOHO) spacecraft for the present solar
cycle (Kosovichev \& Schou 1997; Schou {\em et al.} 1998).

Recent analysis of the helioseismic data, from both
the Michelson Doppler Imager (MDI) instrument on board the SOHO
spacecraft and the Global Oscillation Network Group (GONG)
project, has also produced evidence that this
banding signature is not merely a surface feature, but
extends into the convection zone, to a depth of at least 8 percent in
radius (Howe {\em et al.} 2000).
These authors present data on
departures of the reconstructed rotation rate from its temporal
averages - the residuals - as a function of latitude at several target depths,
which behave in a manner
similar to the migration of sunspots during the solar cycle (the `butterfly
diagram').
This finding is also supported by the analysis of Antia \& Basu (2000),
who use different data sets from GONG and independent inversion techniques.
The time-base of these observations is only a few years, less
than a complete solar cycle.

These torsional oscillations are
thought to be produced as a consequence
of the nonlinear interactions between the magnetic
fields and the solar differential rotation.
A zero order `mean' differential rotation
is assumed to be maintained by the Reynolds stresses of the turbulence;
this can be included as a constant `background' angular velocity, or
explicitly parametrized, e.g.\ by the
so called $\Lambda$--effect (e.g.\ R\"udiger 1989).
Attempts have been made to explain the surface oscillations in terms of the
effects of the Lorentz force exerted by the large scale magnetic field on the
azimuthal velocity field (e.g.\ Brandenburg \& Tuominen 1988), or as a
consequence of the `quenching' of the turbulence-dependent
quantities by the magnetic field
(Kitchatinov {\em et al.} 1999).

Here we study this nonlinear coupling in the context
of a two dimensional axisymmetric mean field
dynamo model, in a spherical shell,
in which the only nonlinearity  is the action
of the azimuthal component of the Lorentz force of the
dynamo generated magnetic field on the solar
angular velocity.
Obtaining torsional oscillations in this
way is also of interest in view
of the fact that the Lorentz force is of second order in
the magnetic field, thus naturally leading to the excitation
of hydrodynamical oscillations of about half the period
of the magnetic oscillations.

In the next section we introduce our model.
Section 3 contains our results and section
4 gives a brief discussion.
\section{The model}
Here we shall assume that the gross features of the
large scale solar magnetic field
can be described by a mean field dynamo
model, with the standard equation
\begin{equation}
\frac{\partial{\bf B}}{\partial t}=\nabla\times({\bf u}\times {\bf B}+\alpha{\bf
B}-\eta\nabla\times{\bf B}),
\label{mfe}
\end{equation}
where ${\bf u}=v\mathbf{\hat\phi}-\frac{1}{2}\nabla\eta$,
the term proportional to
$\nabla\eta$ represents the effects of turbulent diamagnetism,
and where the velocity field is taken to be of the form
\begin{equation}
v=v_0+v',
\end{equation}
where $v_0=\Omega_0 r \sin\theta$, $\Omega_0$ is a prescribed
underlying rotation law and the component $v'$ satisfies
\begin{equation}
\frac{\partial v'}{\partial t}=\frac{(\nabla\times{\bf B})\times{\bf B}}{\mu_0\rho
r \sin\theta} . \mathbf{\hat {\bf \phi}}  + \nu D^2 v',
\label{NS}
\end{equation}
where $D^2$ is the operator
$\frac{\partial^2}{\partial r^2}+\frac{2}{r}\frac{\partial}{\partial r}+\frac{1}
{r^2\sin\theta}(\frac{\partial}{\partial\theta}(\sin\theta\frac{\partial}{\partial
\theta})-\frac{1}{\sin\theta})$ and $\mu_0$ is the induction constant.
The assumption of axisymmetry allows the field ${\bf B}$ to be split into toroidal
and poloidal parts,
${\bf B}={\bf B}_T+{\bf B}_P = B\hat\phi +\nabla\times A\hat\phi$,
and results in Eq.\ (\ref{mfe}) yielding two scalar equations for $A$ and $B$.
Nondimensionalizing in terms of the solar radius $R_{\odot}$ and time $R_{\odot}^2/\eta_0$,
where $\eta_0$ is the maximum value of $\eta$ and
letting $\Omega=\Omega^*\tilde\Omega$, $\alpha=\alpha_0\tilde\alpha$,
$\eta=\eta_0\tilde\eta$, ${\bf B}=B_0\tilde{\bf B}$ and $v'= \Omega^* R_{\odot}\tilde v'$,
results in a system of equations for $A,B$ and $v'$, with the
dynamo parameters $R_\alpha=\alpha_0 R_{\odot}/\eta_0$, $R_\omega=\Omega^*R_{\odot}^2/\eta_0$,
$P_r=\nu_0/\eta_0$, and $\tilde\eta=\eta/\eta_0$, where $\Omega^*$
is the solar surface equatorial angular velocity (see Moss \& Brooke 2000 for details).
Here
$\nu_0$ and $\eta_0$ are the turbulent magnetic
diffusivity and viscosity respectively
and $P_r$ is the turbulent Prandtl number.
The density $\rho$ is assumed to be uniform, and
stress free boundary conditions ensure angular momentum conservation.

These equations were solved using the code
and boundary conditions described in
Moss \& Brooke (2000),
over the range $r_0\leq r\leq1$, $0\leq\theta\leq
\pi$, with uniform spacing in both $r$ and $\theta$.
The computational domain  is the region $r_0=0.64 \leq r \leq 1$; with
the solar convection zone proper being thought to occupy the region $r > 0.7$,
the region $r_0 \leq r \leq 0.7$ can be thought of as an overshoot region/tachocline.
In the following simulations we used a mesh resolution of 61 points uniformly
distributed radially
and 101 points uniformly distributed latitudinally
(over $0 \leq \theta \leq \pi$), but test runs were carried out
at higher spatial resolutions.

In this investigation, we took $\Omega_0$ to be given in
$0.64\leq r \leq 1$ by an interpolation on the MDI data
obtained from 1996 to 1999 (Howe {\em et al.} 2000), depicted in
Fig.\ 1.
For $\alpha$ we took $\tilde\alpha=\alpha_r(r)f(\theta)$,
where
$f(\theta)=\sin^2\theta\cos\theta$
(cf.\ R\"udiger \& Brandenburg 1995)
and
\begin{equation}
\alpha_r=1;~~~~~ 0.7 \leq r \leq 0.8
\end{equation}
with cubic interpolation to zero at $r=r_0$ and $r=1$,
with the convention that $\alpha_r>0$
and $R_\alpha < 0$. Also, in
order to take into account the
likely decrease in the turbulent diffusion coefficient $\eta$
in the overshoot region, we allowed a simple
linear decrease from $\tilde\eta=1$ at $r=0.8$
to $\tilde\eta=0.5$ in $r<0.7$.

We monitor the time evolution of the total magnetic energy $E$
and the global parity of the magnetic field,  defined as
$P=\frac{E^S-E^A}{E^S+E^A}$,
where $S$ and $A$ refer respectively to the parts of the magnetic field that
have symmetry or antisymmetry with respect to the equatorial plane
(see also Brandenburg {\em et al.} 1989).
Thus $P=+1$ and $-1$ correspond to symmetric and antisymmetric
fields respectively.
\section{Results}
We calibrated our model so that near marginal excitation the cycle period was about 22 years.
This determined $R_\omega=6\times 10^4$, corresponding to
$\eta_0\approx 2.5\times 10^{11}$ cm$^2$s$^{-1}$,
given the known values of $\Omega^*$ and $R_{\odot}$.
The first solutions to be excited in the linear theory
are odd parity ($P=-1$) limit cycles, which in this case
have marginal dynamo number
$R_\alpha \sim -3.12$. The even parity ($P=+1$) solutions
are also excited at similar marginal dynamo numbers of  $R_\alpha \sim -3.16$.
We considered two values of the Prandtl numbers,
$P_r = 0.1$ and $P_r = 1.0$.
\begin{figure}[!htb]
\centerline{\def\epsfsize#1#2{0.5#1}\epsffile{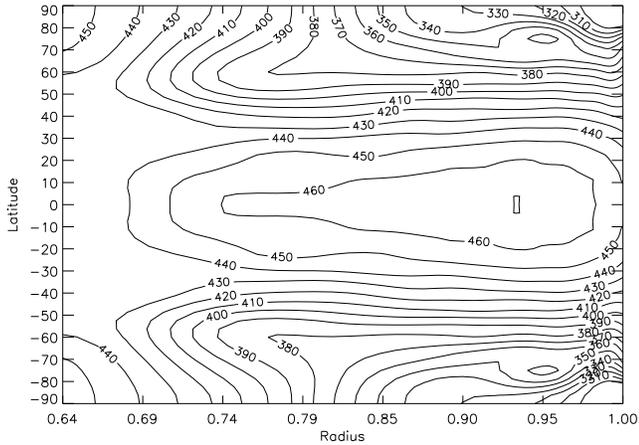}}
\caption{\label{mdi} Isolines of the time average of the angular velocity of the
solar rotation, obtained by inverse techniques
using the MDI data (Howe {\em et al.} 2000).
Contours are labelled in units of nHz.}
\end{figure}

\begin{figure}[!htb]
%
%
\centerline{\def\epsfsize#1#2{0.42#1}\epsffile{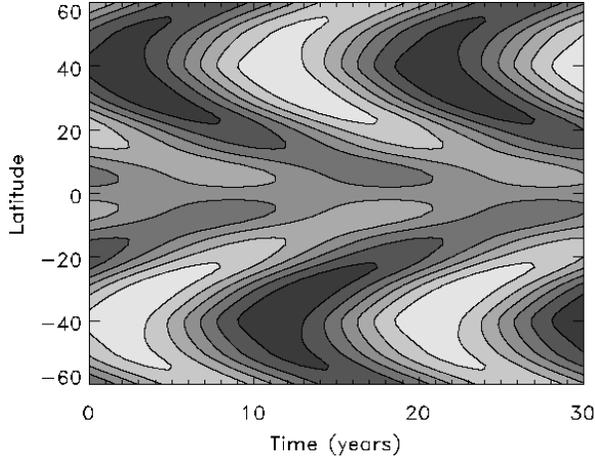}}
\caption{\label{Calph=-3.2.Pr=1.0.MDI2.toroidal.butterfly}
Butterfly diagram of the toroidal component of the
magnetic field $\vec{B}$ at $R=0.95 R_{\odot}$.
Dark and light shadows correspond to positive and negative
$B_\phi$ respectively.
Parameters values
are $R_{\alpha}=-3.2$, $P_r=1.0$ and ${R_{\omega}}=6\times 10^{4}$.
}
\end{figure}

With these parameter values, we found that this model,
with underlying zero order angular velocity
chosen to be consistent with the recent (MDI) helioseismic
data (Fig.\ \ref{mdi}), is capable of
producing butterfly diagrams which are in qualitative agreement
with the observations. An example of this is depicted in
Fig.\ \ref{Calph=-3.2.Pr=1.0.MDI2.toroidal.butterfly}.
The polar feature is rather too strong -- we have checked that
this can be rectified
by modifying slightly the spatial dependence of $\alpha$, by for example
choosing $f(\theta)=\sin^4\theta\cos\theta$.

\begin{figure}[!htb]
\centerline{\def\epsfsize#1#2{0.42#1}\epsffile{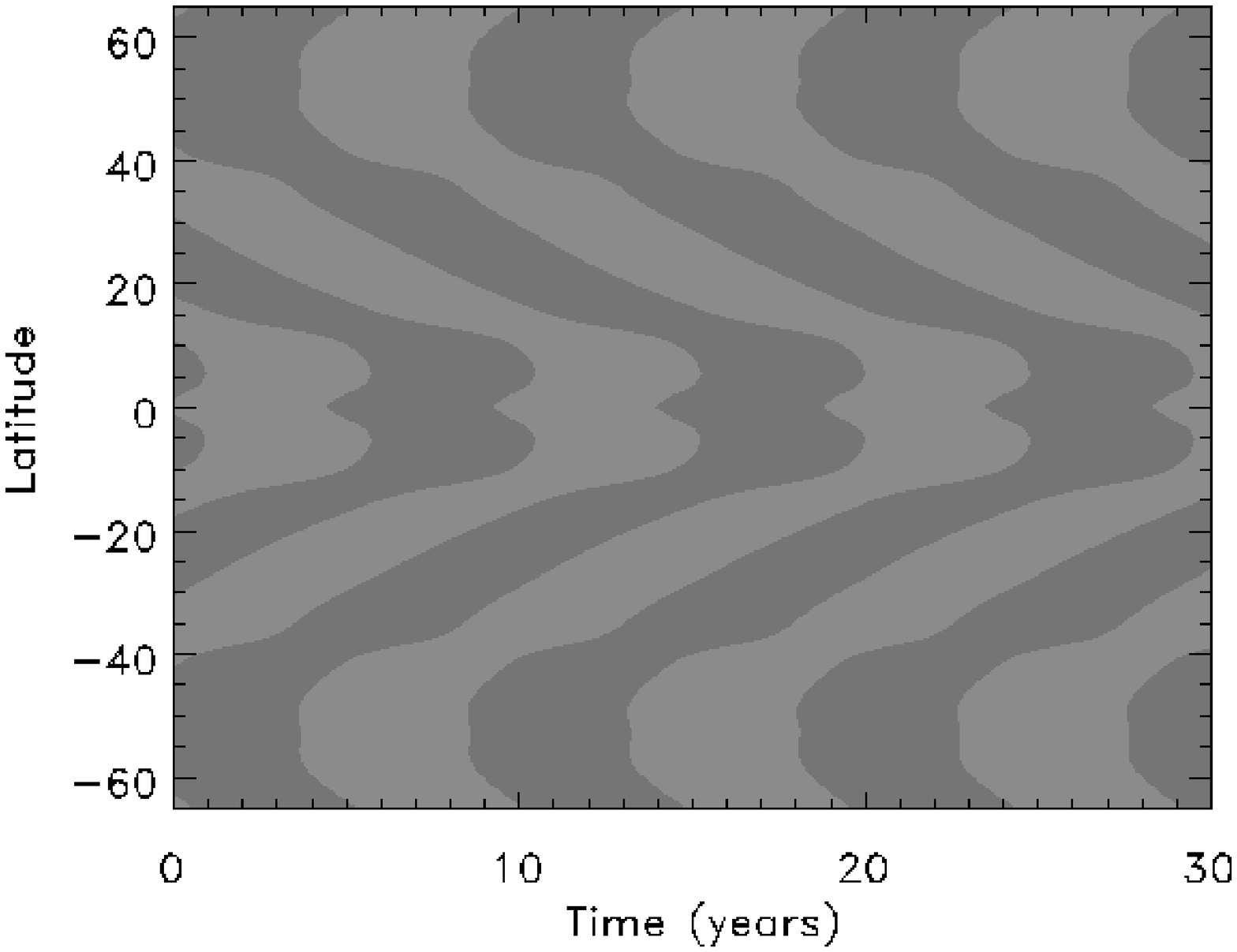}}
\centerline{\def\epsfsize#1#2{0.42#1}\epsffile{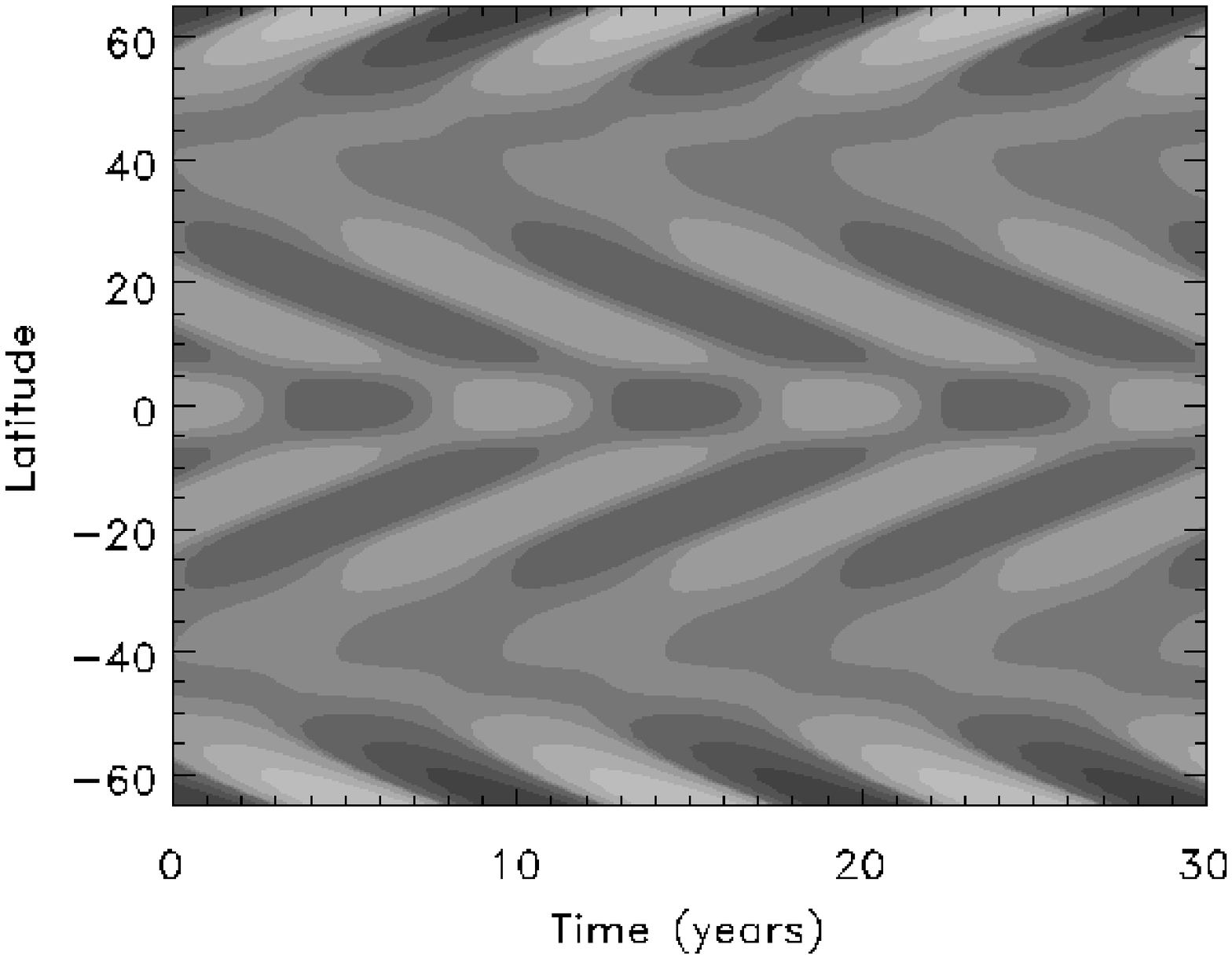}}
\caption{\label{Calph=-3.2.Pr=0.1.MDI2.velocity}
Variation of rotation rate with latitude and time from
which a temporal average has been subtracted to reveal the
migrating banded zonal flows, at $R=0.95  R_{\odot}$ (top) and
$R=0.84  R_{\odot}$ (bottom). Darker and lighter regions
represent positive and negative deviations
from the time averaged background rotation rate.
Parameters values
are $R_{\alpha}=-3.2$, $P_r=0.1$ and ${R_{\omega}}=6\times 10^{4}$. }
\end{figure}

\begin{figure}[!htb]
\centerline{\def\epsfsize#1#2{0.42#1}\epsffile{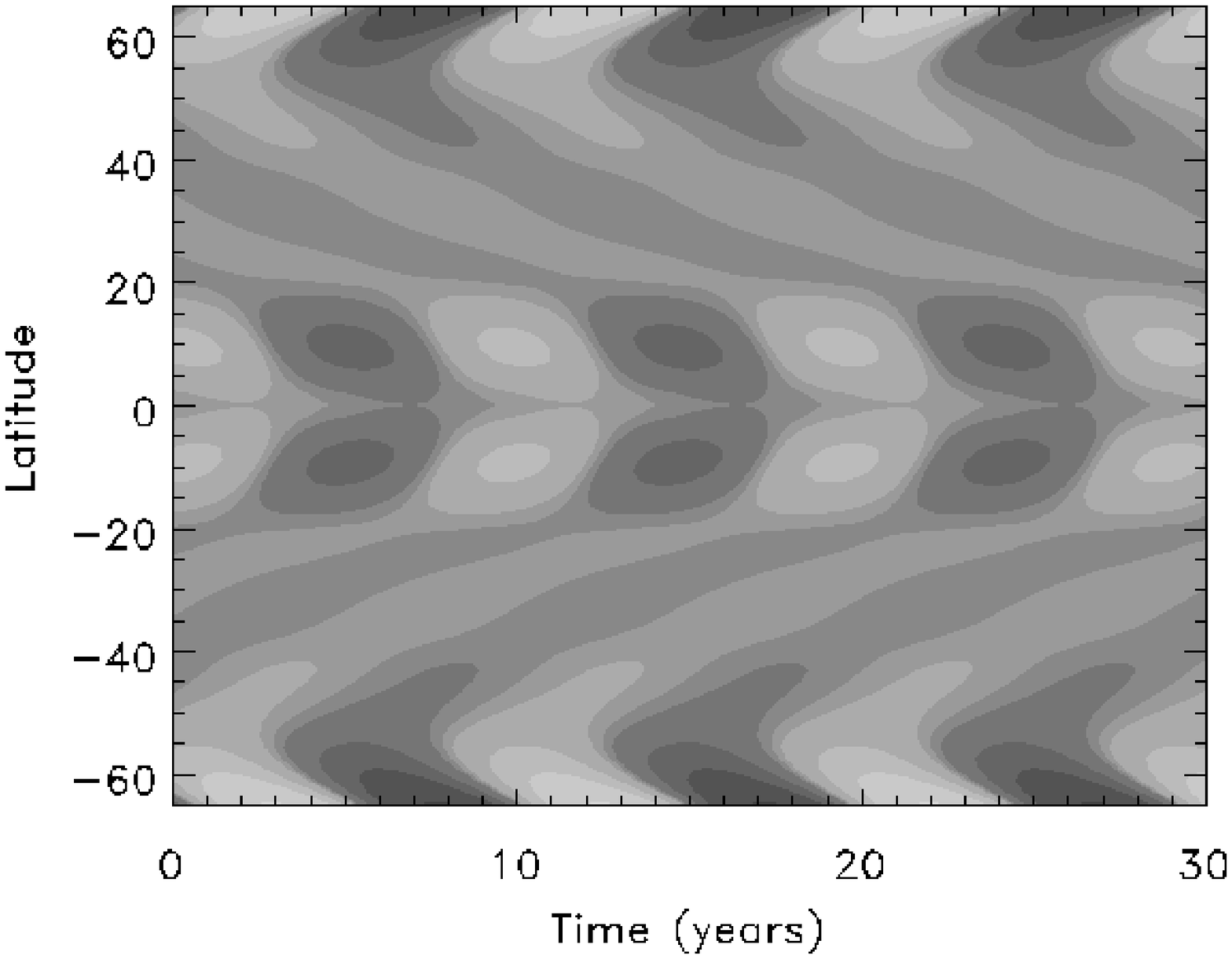}}
\centerline{\def\epsfsize#1#2{0.42#1}\epsffile{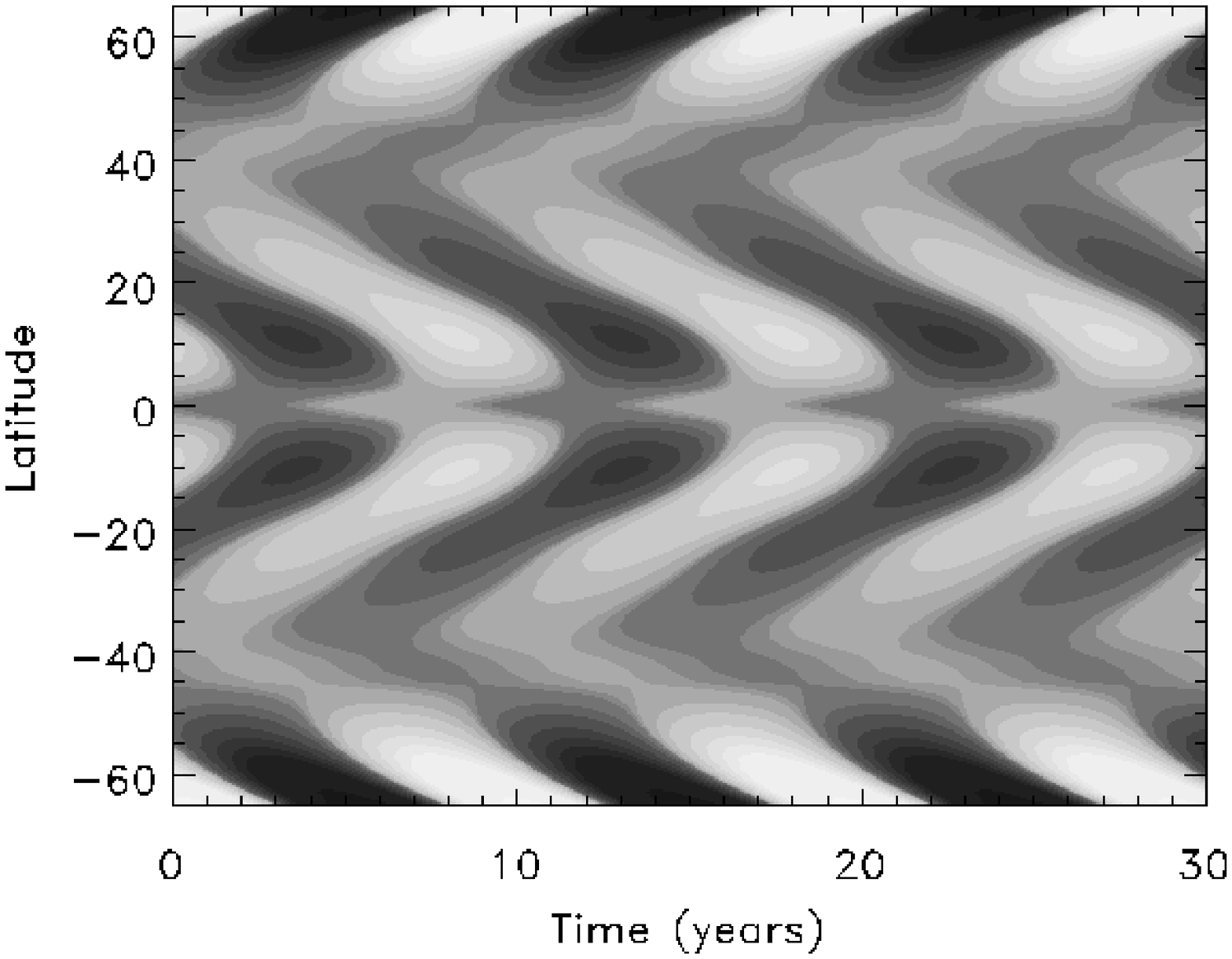}}
\caption{\label{Calph=-3.2.Pr=1.0.MDI2.velocity}
Variation of rotation rate with latitude and time from
which a temporal average has been subtracted to reveal the
migrating banded zonal flows, at $R=0.95  R_{\odot}$ (top) and
$R=0.84  R_{\odot}$ (bottom).
Parameters values
are $R_{\alpha}=-3.2$, $P_r=1.0$ and ${R_{\omega}}=6\times 10^{4}$. }
\end{figure}

We also found that this model successfully produced torsional oscillations
in the convection zone, similar to those deduced from
recent helioseismic data. To compare our model with these
results, we have plotted in Figs.\ \ref{Calph=-3.2.Pr=0.1.MDI2.velocity} and
\ref{Calph=-3.2.Pr=1.0.MDI2.velocity} the variations of the rotation rate
with latitude and time (`butterfly diagrams')
to reveal the
migrating banded zonal flows.
In these models, the basic magnetic field parity is odd ($P=-1$).

\begin{figure}[!htb]
\centerline{\def\epsfsize#1#2{0.40#1}\epsffile{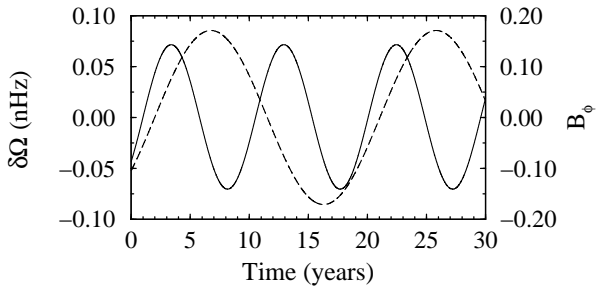}}
\caption{\label{Calph=-3.2.Pr=1.0.MDI2.time.series.residuals}
Variation of rotation rate at radius $0.84 R_{\odot}$ and
latitude $60$ degrees. Also depicted is the
(nondimensionalized) toroidal field at the same radius and latitude
(dashed line), which shows double the period. Parameters values
are as in Fig. 4.
}
\end{figure}

For the sake of comparison with observational results,
we have also plotted in Fig.\ \ref{Calph=-3.2.Pr=1.0.MDI2.time.series.residuals}
the evolution of the residual
rotation rate with time, at radius $0.84$
and latitude $60$ degrees.

As can be seen,  consistent with the observations,
in each hemisphere there are
alternating latitudinal bands, with the width of approximately
10 degrees, of slightly faster and slower than average zonal flows.
These  migrate towards the equator in about 22
years, and
extend deep into the
convection zone. The amplitudes of these oscillations
increase with depth below the
surface and depend on the parameters of our
model, in particular the Prandtl number.
For $P_r = 1.0$, these amplitudes range from
about 0.07 nHz at the surface to more than $0.4$ nHz
towards the bottom of the convection zone,
somewhat lower than, but in principle compatible with,
the results of Howe {\em et al.} 2000.
The torsional
oscillations present in our model have periods
half that of the period of
the global magnetic field which is
compatible with the observed period
of the oscillations at the surface
and consistent with the observed behaviour
inside the convection zone.

The strictly odd parity models presented have an equatorial feature in the
`butterfly diagrams' for the velocity perturbations which does not
appear to be present in the current inversions of the observational data.
This becomes weaker as Prandtl number increases. We also note in
passing that these figures are almost identical for cases where even parity
solutions are found, except that this feature is then absent. We further point
out that the large scale solar magnetic field is probably of {\it mixed}
global parity (predominantly odd) -- see, e.g., Pulkinnen {\em et al.} (1999).
\section{Discussion}
We have studied a solar dynamo model, calibrated to have the correct
cycle period, with a mean rotation law given by recent
helioseismic observations. This
model produces butterfly diagrams in qualitative
agreements with those observed.

We have shown that a nonlinear coupling between
the magnetic field and the solar differential rotation,
where the nonlinearity is due to the action
of the azimuthal component of the Lorentz force,
is capable of producing torsional oscillations,
with period of about 11 years,
which penetrate into the convection zone
and which migrate towards the equator in about
22 years. The period of these
oscillations is about half that of the period of
the global magnetic fields. This is in agreement with
the observed period of the torsional oscillations at the surface.
For the oscillations inside the convection zone,
this is a testable prediction, not contradicted by the
current helioseismic
observations, which so far extend over an interval less than
a solar cycle.
We note also that solutions with even parity ($P =+1$)
show slightly larger amplitudes without the equatorial feature.

The current inversions of the helioseismological data
seem to suggest that the torsional
oscillations largely disappear
below about $R =0.9 R_{\odot}$, in contrast
to our model oscillations.
However, there are uncertainties in these inversions, specially
at the deeper levels. At the same time
our dynamical model is oversimplified and
substantial improvements can be made.
In particular, the predicted amplitudes for torsional
oscillations in our model are likely to be affected
by our assumption of uniform density in Eq.\ (\ref{NS}).
Nevertheless,
we find it interesting that some of
the major features in the torsional
oscillations can be readily reproduced.
A more detailed study of these
oscillations, including more
realistic density profiles,
is in progress.
\begin{acknowledgements}
We would like to thank Rachel Howe and Mike Thompson for
many useful discussions regarding the helioseismic observations
and for providing us with the data displayed in Fig.\ \ref{mdi}.
EC is supported by a PPARC fellowship and RT
benefited from PPARC UK Grant No.\ L39094.
\end{acknowledgements}

\end{document}